\documentclass[prb,twocolumn,showpacs,floatfix,amsmath,amssymb,superscriptaddress]{revtex4}
\usepackage[dvips]{graphicx}
\usepackage{latexsym}
\usepackage{graphicx}
\usepackage{times}
\usepackage{amsmath}
\usepackage{dcolumn}
\usepackage{subfigure}
\usepackage{latexsym,amsmath,amssymb,bm,euscript}
\usepackage{amsfonts}
\usepackage{hyperref}

\usepackage{color}

\def\etal{~\textit{et~al.}} % etal 

\newcommand{\ket}[1]{| #1 \rangle}
\newcommand{\bra}[1]{\langle #1 |}

\newcommand{\be}{\begin{equation}}
\newcommand{\ee}{\end{equation}}

\newcommand{\eps}{\epsilon}

\begin{document}

\title{Realization of a vortex in the Kekule texture of molecular Graphene, at a Y junction where 3 domains meet}  
\author{Doron L. Bergman}
\affiliation{Physics Department, California Institute of Technology,
  MC 114-36, 1200 E. California Blvd., Pasadena, CA 91125}

\date{\today} 
 
\begin{abstract}
Following the recent realization of an artificial version of Graphene in the electronic surface states of copper with judiciously placed carbon monoxide molecules inducing the honeycomb lattice symmetry (K. K. Gomes \etal Nature 483, 306 (2012)),  
we demonstrate that these can be used to realize a vortex in a Kekule texture of the honeycomb lattice.  The Kekule texture is mathematically analogous to a superconducting order parameter, opening a spectral gap in the massless Dirac point spectrum of the Graphene structure. The
core of a vortex in the texture order parameter, supports subgap states, which for this system
are analogs of Majorana fermions in some superconducting states. In particular, the electron charge bound to a single vortex core is effectively fractionalized to a charge of $e/2$.
The Kekule texture as realized in the molecular Graphene system realizes 3 different domain types, and we show that a Y-junction between them realizes the coveted Kekule vortex.
%We further discuss (cite Caroli DeGennes Matricone)
\end{abstract} 
%\pacs{75.10.Jm, 75.40.Gb} 
 
\maketitle 

%\section{Introduction}

The experimental realization of Graphene\cite{Novoselov:2005}, has inspired in the last few years a large body of work exploring the possible physics in an ideal 2D system, realizing massless Dirac fermions, including a great deal of theoretical work\cite{Graphene_RMP}. While Graphene is proving a very flexible medium to manipulate, as a physical system it has its limitations, and sadly, some of the most interesting physical effects theoretical work suggested in Graphene-like systems have not been realized in the actual Graphene system. For this reason, there is good reason to explore alternative realizations of the single layer honeycomb 2D electron gas.

\begin{figure}[htbp]
\begin{center}
\includegraphics[width=3.0in]{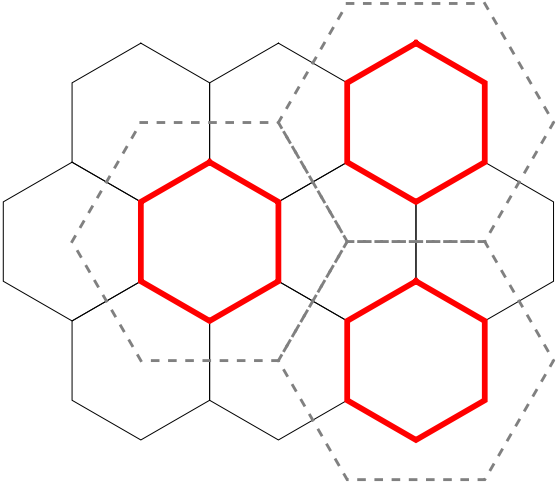}
\caption{Kekule pattern on the honeycomb lattice. The thick (red) links have stronger (or weaker) hopping strength on them. The unit cell is denoted by dashed (gray) lines. }
\label{fig:kekule_pattern}
\end{center}
\end{figure}

Recent advances in STM technology have allowed to manufacture an artificial version Graphene, by arranging carbon monoxide molecules on the surface of copper, dubbed "molecular Graphene".
The carbon monoxide molecules are arranged in a regular array, and thus create an electrostatic potential with minima forming the vertices of a honeycomb lattice\cite{Gomes:2012,Simon:2012}. Any 2D electron gas with the symmetry of the honeycomb lattice imposed on it is likely to realize an analog of Graphene. Because of its microscopic construction, the molecular Graphene system is even more easily manipulated than Graphene. In particular, since the electrostatic potential is essentially under full control by selecting an appropriate molecule arrangement, the honeycomb lattice can be engineered with a wide variety of lattice textures, which are predicted to realize the analog of huge magnetic fields\cite{Guinea:2010,Levy:science2010,Guinea:prb2010,Guinea:prb2008,Pereira:prl2009}, as well as analogs of superconducting states\cite{Hou:prl2007,Seradjeh:prl2008,Bergman:prb2009,Herbut:prb2010}. Among the most intriguing of these proposals is the suggestion to realize a Kekule texture
on the honeycomb lattice\cite{Hou:prl2007,Hou:prb2008,Hou:prb2008b}. The Kekule texture makes some nearest neighbor links on the honeycomb lattice stronger than others, in a $\sqrt{3} \times \sqrt{3}$ arrangement as depicted in Fig.~\ref{fig:kekule_pattern}. In the low energy effective massless Dirac Hamiltonian for the honeycomb tight binding model, this arrangement induces a pairing-like term between the electrons of one Dirac point and the holes of the second Dirac point (instead of electrons with opposite spin). In principle the Kekule order parameter can have a complex value, and can include a vortex. It has been shown\cite{Hou:prl2007} that this vortex is the analog of a vortex in a $p_x+ip_y$ superconducting state, supporting a zero mode at the vortex core\cite{Read:prb2000}. In analogy to some systems proposed in the context of high energy physics\cite{Jackiw1981681,Jackiw:1976}, and other solid state systems\cite{Fu:prl2008,Seradjeh:prb2008b} the vortex core is expected to bind a fermion number of one half, in the Kekule texture case - half an electron, if spin is ignored. The halving of a fermion is also at the heart of the emergence of Majorana fermions in superconducting states - there the fermion being halved is the Bogoliubov quasiparticle. The halved electron realized by the Kekule vortex would be a mathematical analog of a Majorana fermion. In this paper we demonstrate how a Kekule vortex can be realized in the molecular Graphene system, opening up the possibility to explore directly the physics of fermion halving.

At the microscopic level a uniform Kekule texture enlarges the unit cell of the honeycomb lattice 3 fold (as illustrated in Fig.~\ref{fig:kekule_pattern}). The unit cell includes three plaquettes, one of which has the nearest neighbor hopping strength on the links around it stronger (or weaker). With any choice of unit cell there are three choices where to locate the Kekule distorted plaquettes. We will show that the Y-junction between these three domains realizes a vortex in the Kekule texture, binding a charge $e/2$ (per spin) to its core.

%\section{Calculation}

Even without the texture, we can use the same enlarged unit cell to describe the Graphene band structure, from which we can identify the low energy massless Dirac Hamiltonian. Afterwards, we add the Kekule texture.

\begin{figure}[htbp]
\begin{center}
\includegraphics[width=2.0in]{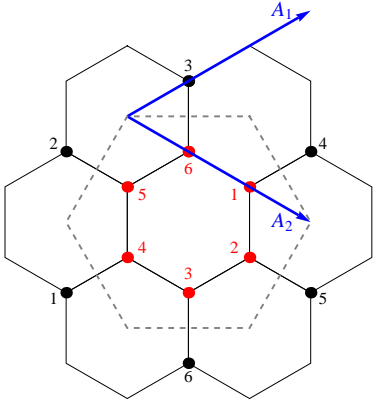}
\caption{The unit cell of the Kekule pattern. The six sites in the unit cell are denoted by $1 \ldots 6$ (red). The unit cell is denoted by the dashed (gray) line. The (blue) Bravais lattice vectors are denoted by ${\bf A}_{1,2}$.}
\label{fig:unit_cell}
\end{center}
\end{figure}

The enlarged unit cell, as depicted in Fig.~\ref{fig:unit_cell} has 6 lattice sites, 
instead of 2 in the original unit cell.
We therefore denote the six sites in the unit cell by $\mu = 1 \ldots 6$,
at positions ${\bf a}_{\mu} = \left(
\sin\left( \frac{2 \pi \mu}{6} \right),
\cos\left( \frac{2 \pi \mu}{6} \right) 
\right)$ relative to the unit cell center.
The original Bravais lattice basis of ${\bf a}_1 + {\bf a}_2$ and
${\bf a}_5 + {\bf a}_6$ is replaced with
${\bf A}_1 = 3 {\bf a}_1$
and
${\bf A}_2 = 3 {\bf a}_2$.
Finally, the enlarged real space unit cell, corresponds in momentum space to copying the band structure with a shift of the reciprocal lattice vectors
${\bf B}_1 = \frac{2\pi}{3} \left( \frac{1}{\sqrt{3}},1 \right)$
and 
${\bf B}_2 =  \frac{2\pi}{3} \left( \frac{1}{\sqrt{3}},-1 \right)$. The first Brillouin Zone is 3 times smaller,
and both Dirac points are shifted to ${\bf q} = 0$, since the Dirac points appear
at ${\bf Q} = \pm ({\bf B}_1 + {\bf B}_2)$.

The tight binding Hamiltonian is
\be\label{Hamiltonian}
\begin{split}
{\mathcal H}_0 = t \sum_{\bf r} \sum_{\mu=1}^6
c_{\mu}^{\dagger}({\bf r}) 
\left[
c_{\mu+1}^{\phantom\dagger}({\bf r})
+ c_{\mu-1}^{\phantom\dagger}({\bf r})
+ c_{\mu+3}^{\phantom\dagger}({\bf r} + {\bf A}_{\mu})
\right]
\; ,
\end{split}
\ee
where ${\bf r}$ denotes the unit cell, ${\bf A}_{\mu} = 3 {\bf a}_{\mu}$ are
Bravais lattice vectors, and the $\mu \pm 1,\mu+3$ indices are added modulo $6$.
The hopping strength is set to $t=1$ for convenience.
Going to momentum space via the Fourier transform
$
c_{\mu}^{\phantom\dagger}({\bf r}) 
= \int d^2{\bf q} \, e^{i ({\bf r  + a}_{\mu}) \cdot {\bf q} } 
c_{\mu}^{\phantom\dagger}({\bf q})
$ ,
the Hamiltonian becomes
\be
{\mathcal H}_0 = \int d^2{\bf q} \; 
c_{\mu}^{\dagger}({\bf q})
H_0^{\mu \nu}({\bf q})
c_{\nu}^{\phantom\dagger}({\bf q})
\; ,
\ee
where 
\be
H_0^{\mu \nu}({\bf q}) = 
e^{i {\bf q} \cdot \left( {\bf a}_{\nu} - {\bf a}_{\mu} \right)} 
\left( \delta_{\mu+1,\nu} + \delta_{\mu-1,\nu} \right)
+
e^{i {\bf q} \cdot {\bf a}_{\mu}} \delta_{\mu+3,\nu}
\; .
\ee
The eigenstates are found from the equation
$
H_0({\bf q}) \psi({\bf q}) = \eps({\bf q}) \psi({\bf q})
$.

The key to identifying the Dirac points in this matrix form is to perform an 
appropriate basis change. We know that we could have chosen a smaller unit cell 
with just 2 sites (wavefunction weights $\chi_{1,2}$), and a choice of 
%$
%\psi({\bf q}) = \left( 
%\psi_1, \,
%\psi_2, \,
%\psi_1 e^{i {\bf q} \cdot ({\bf a}_3 - {\bf a}_1)}, \,
%\psi_2 e^{i {\bf q} \cdot ({\bf a}_4 - {\bf a}_2)}, \,
%\psi_1 e^{i {\bf q} \cdot ({\bf a}_4 - {\bf a}_1)}, \,
%\psi_2 e^{i {\bf q} \cdot ({\bf a}_6 - {\bf a}_2)}
%\right)^T
%$,
$
\psi({\bf q}) = \left( 
\chi_1,
\chi_2,
\chi_1 e^{i {\bf q} \cdot {\bf a}_{3,1}},
\chi_2 e^{i {\bf q} \cdot {\bf a}_{4,2}},
\chi_1 e^{i {\bf q} \cdot {\bf a}_{4,1}},
\chi_2 e^{i {\bf q} \cdot {\bf a}_{6,2}}
\right)^T
$,
where ${\bf q} = 0, \pm {\bf B}_1$, and ${\bf a}_{i,j} = {\bf a}_i - {\bf a}_j$
should recover that choice, since it recovers the plane wave phases
between the different sites in the unit cell, while keeping the 2 sites of the small unit cell
unchanged.
This therefore suggests using the unitary transformation comprised of
$
U = \left( 
\psi_1(0) ,
\psi_2(0) ,
\psi_1({\bf B}_1) ,
\psi_2({\bf B}_1) ,
\psi_1(-{\bf B}_1) ,
\psi_2(-{\bf B}_1) 
\right)
$ ,
where
$
\psi_1({\bf q}) = \left( 
1,
0, 
e^{i {\bf q} \cdot ({\bf a}_3 - {\bf a}_1)},
0,
e^{i {\bf q} \cdot  ({\bf a}_4 - {\bf a}_1)}, 
0
\right)^T
$,
and
$
\psi_2({\bf q}) = \left( 
0,
1, 
0, 
e^{i {\bf q} \cdot  ({\bf a}_4 - {\bf a}_2)},
0,
e^{i {\bf q} \cdot  ({\bf a}_6 - {\bf a}_2)}
\right)
$.
This turns out to be
\be\label{unitary}
U = 
 \frac{1}{\sqrt{3} }
\left(
\begin{array}{cccccc}
 1 & 0 & 1 & 0 & 1 & 0 \\
 0 & 1 & 0 & 1 & 0 & 1 \\
 1 & 0 & \omega& 0 & \omega^2 & 0 \\
 0 & 1 & 0 & \omega^2 & 0 & \omega \\
 1 & 0 & \omega^2 & 0 & \omega & 0 \\
 0 & 1 & 0 & \omega & 0 & \omega^2
\end{array}
\right)
\; ,
\ee
where $\omega = e^{i 2\pi/3}$.
Using this unitary transformation, expanding $H_0$ to linear order in ${\bf q}$ 
and taking $t = \frac{2}{3}$, we find
\be
U^{\dagger} H_0 U = 
\left(
\begin{array}{cccccc}
 0 & 2 & 0 & 0 & 0 & 0 \\
 2 & 0 & 0 & 0 & 0 & 0 \\
 0 & 0 & 0 & q_- & 0 & 0 \\
 0 & 0 & q_+ & 0 & 0 & 0 \\
 0 & 0 & 0 & 0 & 0 & -q_+ \\
 0 & 0 & 0 & 0 & -q_- & 0
\end{array}
\right)
\; ,
\ee
where $q_{\pm} = q_1 \pm i q_2$. 
The structure of the two Dirac points is now easily seen in the diagonal blocks
in the third to sixth columns. The diagonal $2 \times 2$ block in the first and second columns
corresponds to high energy modes we will ignore for the low energy theory.

\begin{figure}[htb]
\centering
\subfigure[Choice $A$]{
\includegraphics[scale=0.3]{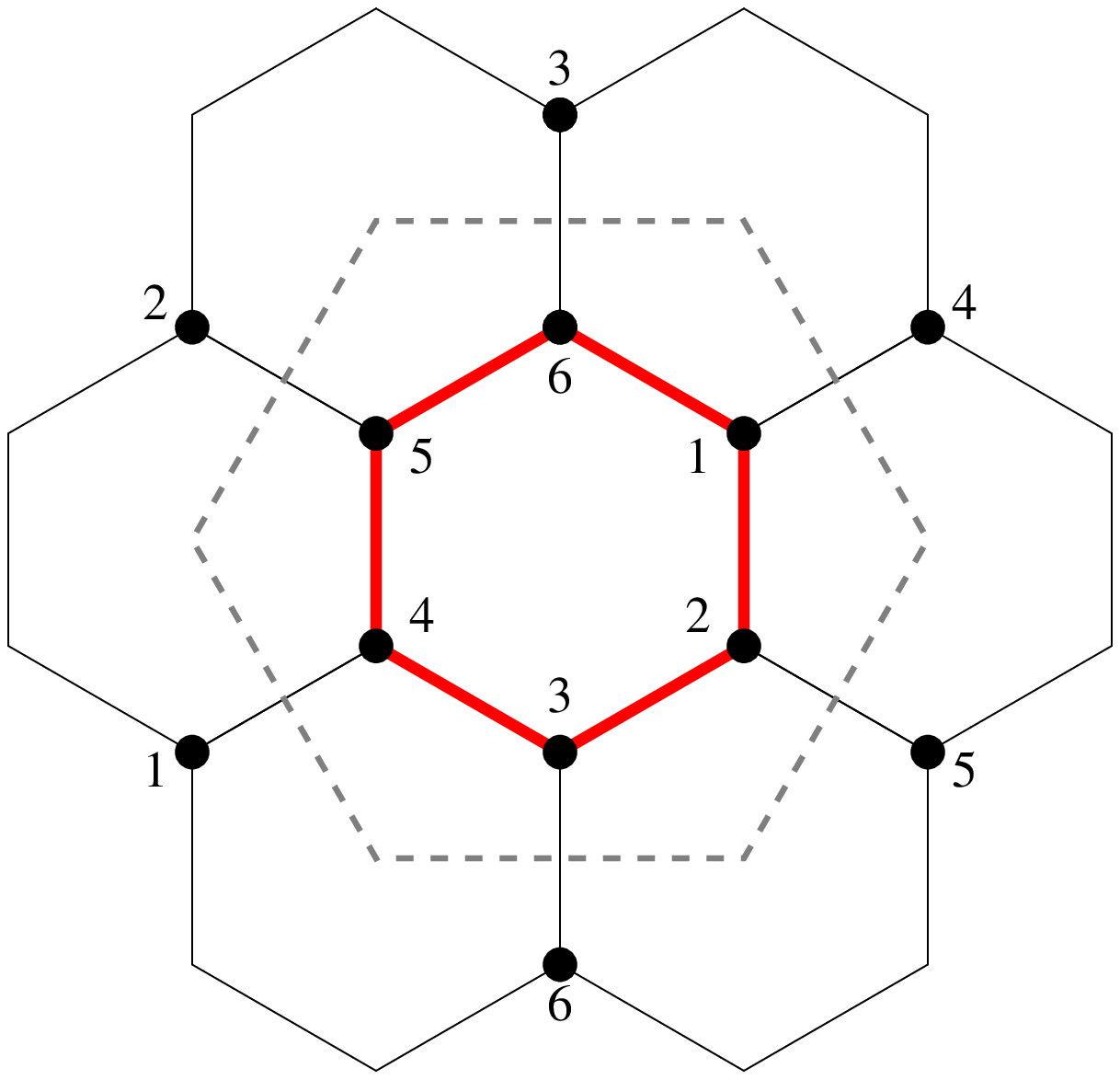}
\label{fig:domain1}
}
\subfigure[Choice $B$]{
\includegraphics[scale=0.3]{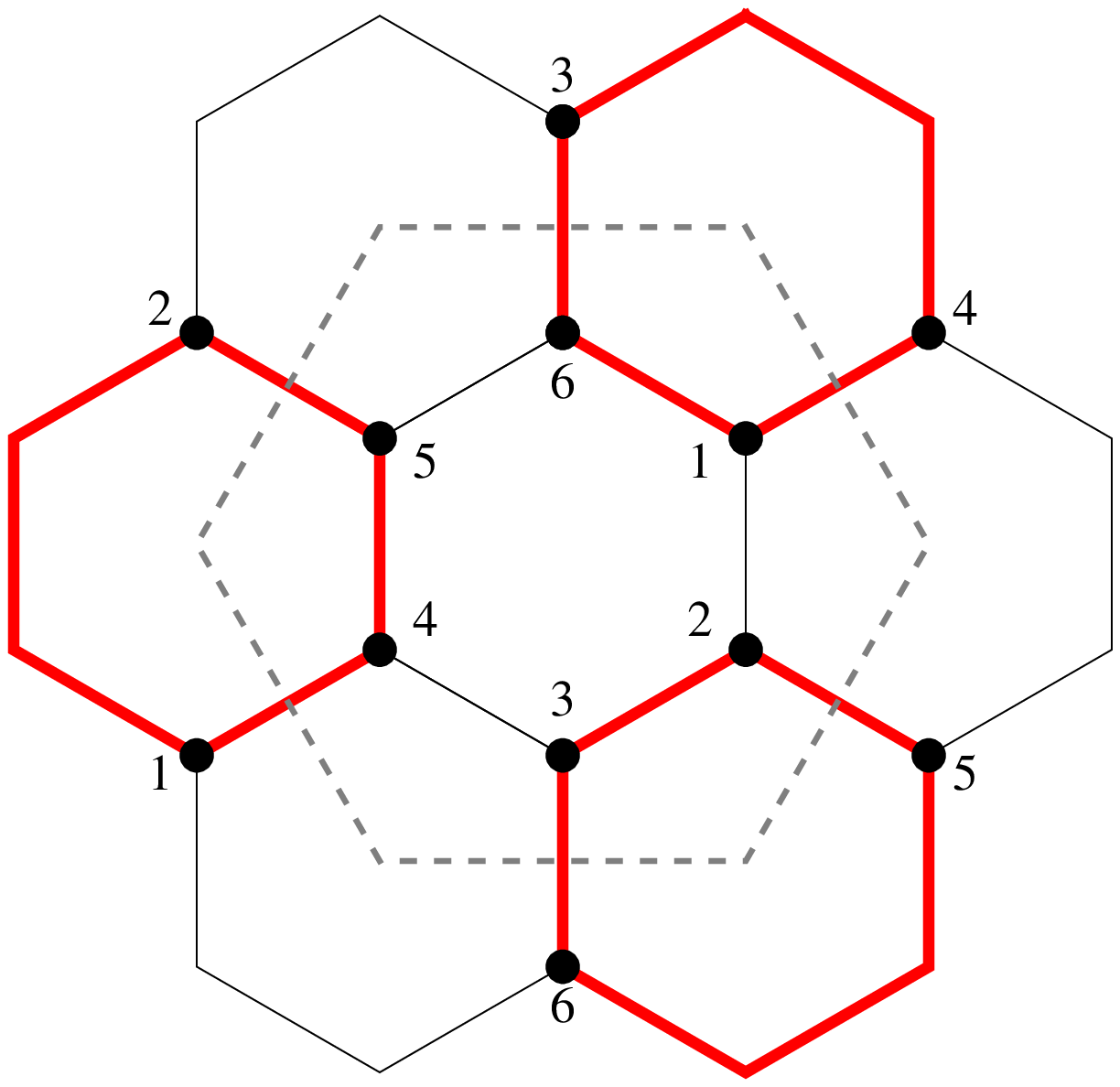}
\label{fig:domain2}
}
\subfigure[choice $C$]{
\includegraphics[scale=0.3]{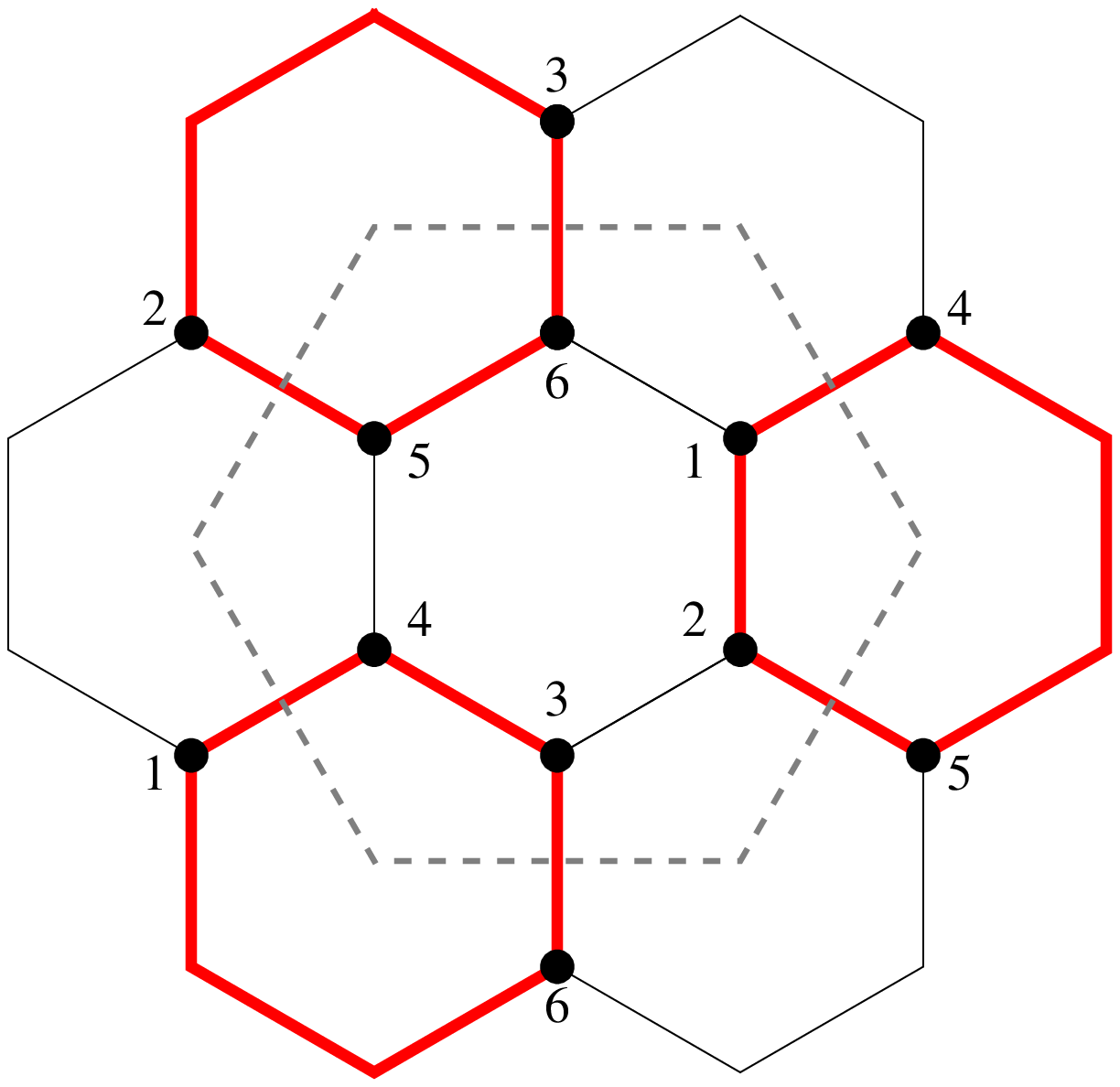}
\label{fig:domain3}
}
\caption{The three different Kekule domains, with a specific choice of unit cell.
These three domains are characterized by an effective phase difference of $2\pi/3$.}
\label{fig:domains}
\end{figure}

Next we add the Kekule texture to the Hamiltonian, with strength $\lambda$.
With any choice of unit cell there are three choices of the Kekule pattern,
shown in Fig.~\ref{fig:domains}. With our choice of unit cell the additional hopping strength in each case can be quantified by adding to $H_0$ the terms $H_{\lambda,\alpha=1,2,3}$,
where
\be
\begin{split} &
H_{\lambda,1}^{\mu \nu}({\bf q}) =
\\ &
\lambda e^{i {\bf q} \cdot \left( {\bf a}_{\nu} - {\bf a}_{\mu} \right)} 
\left( \delta_{\mu+1,\nu} + \delta_{\mu-1,\nu} \right)
\\ &
H_{\lambda,2}^{\mu \nu}({\bf q}) = 
\\ &
\lambda \left[
e^{i {\bf q} \cdot \left( {\bf a}_{\nu} - {\bf a}_{\mu} \right)} 
\left( \delta_{\mu+1,\nu} \delta_{\mu,\text{even}} + \delta_{\mu-1,\nu} \delta_{\mu,\text{odd}} \right)
+
e^{i {\bf q} \cdot {\bf a}_{\mu}} \delta_{\mu+3,\nu}
\right]
\\ & H_{\lambda,3}^{\mu \nu}({\bf q}) =
\\ & 
\lambda \left[
e^{i {\bf q} \cdot \left( {\bf a}_{\nu} - {\bf a}_{\mu} \right)} 
\left( \delta_{\mu+1,\nu} \delta_{\mu,\text{odd}} + \delta_{\mu-1,\nu} \delta_{\mu,\text{even}} \right)
+
e^{i {\bf q} \cdot {\bf a}_{\mu}} \delta_{\mu+3,\nu}
\right]
\; ,
\end{split}
\ee
corresponding to the patterns in Fig.~\ref{fig:domain1},
Fig.~\ref{fig:domain2} and Fig.~\ref{fig:domain3}
respectively.

Taking now $H = H_0 + H_{\lambda}$ for the three cases, expanding to linear order in small 
${\bf q}$, and finally applying the unitary transformation \eqref{unitary}, we find for $H_{\lambda,1}$
\be
\begin{split} &
U^{\dagger} \left(H_0 + H_{\lambda,1}\right) U =
\\ &
\left(
\begin{array}{cccccc}
0 & 2{\tilde \lambda} & 0 & \frac{\lambda}{2}\omega q_+ & 0 & -\frac{\lambda}{2}\omega^2 q_- \\
2{\tilde \lambda} & 0 & \frac{\lambda}{2} \omega q_- & 0 & -\frac{\lambda}{2}\omega^2 q_+ & 0 \\
0 & \frac{\lambda}{2}\omega^2 q_+ & 0 & {\tilde \lambda}q_- & 0 & -\lambda \omega \\
\frac{\lambda}{2}\omega^2 q_- & 0 & {\tilde \lambda}q_+ & 0 & -\lambda \omega & 0 \\
0 & -\frac{\lambda}{2}\omega q_- & 0 & -\lambda \omega^2 & 0 & -{\tilde \lambda}q_+ \\
-\frac{\lambda}{2}\omega q_+ & 0 & -\lambda \omega^2 & 0 & -{\tilde \lambda}q_- & 0
\end{array}
\right)
\; ,
\end{split}
\ee
where ${\tilde \lambda} = 1+\lambda$.
We project out the high energy modes involving the first and second columns and rows, and retain only the low energy Hamiltonian blocks, which yield
\be
\begin{split} &
U^{\dagger} \left(H_0 + H_{\lambda,\alpha=1,2,3}\right) U =
\\ &
\left(
\begin{array}{cccc}
 0 & {\tilde \lambda} q_- & 0 & -\lambda \omega^{\alpha}  \\
 {\tilde \lambda} q_+ & 0 & -\lambda \omega^{\alpha}  & 0 \\
 0 & -\lambda \omega^{2\alpha} & 0 & -{\tilde \lambda} q_+ \\
 -\lambda \omega^{2\alpha} & 0 & -{\tilde \lambda} q_- & 0
\end{array}
\right)
\; .
\end{split}
\ee
We find that the Kekule texture renormalizes the Fermi velocity 
by ${\tilde \lambda} = 1+\lambda$, and otherwise gives a term mixing between the two Dirac cones.
The mixing term has a different phase for the three domains, giving a Kekule texture order parameter $\Delta = \lambda \omega^{\alpha}$, where $\alpha = 1,2,3$. The relative differences in phase being $\pm \frac{2\pi}{3}$.
From this fact we conclude that at a Y junction between the three different domains of Kekule texture, there will be a phase winding of $2\pi$, thus realizing a vortex.

\begin{figure}
	\centering
\includegraphics[scale=0.4]{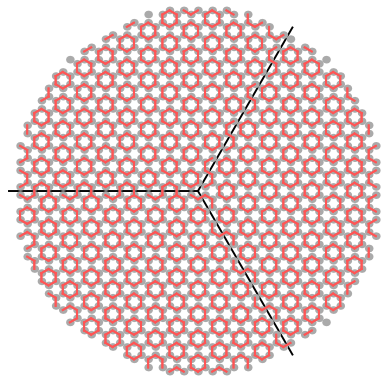}
	\caption{Image of the Kekule texture Y--junction, which we use in numerics. The Graphene flake is a disc of radius $22.5$, using the convention mentioned in the text. The stronger hopping links are denoted by a thick (red) link between the (gray) points denoting the lattice sites. The boundaries between the three Kekule texture domains are denoted by (black) lines, indicating the Y--junction shape. The point where the three domain walls meet, is a vortex core. An effective phase jumps by $2\pi/3$ at each domain wall.}
	\label{fig:Kekul_vortex}
\end{figure}

%\section{Numerics}

Next we present a numerical calculation of the local density of states (LDOS)
near the Kekule vortex.
The LDOS is a particularly useful quantity in the context of molecular Graphene since STM is used to construct the molecular Graphene to begin with, and the same STM can be used to measure the LDOS. We will use this to try and probe the unique states bound to the Kekule vortex.
We use the tight binding model ${\mathcal H}_0$ on a finite, disc-shaped flake of Graphene (of radius $22.5$ in the convention we use here), 
with Kekule textures added realizing the Y-junction between the three domains of the Kekule texture, as shown in Fig.~\ref{fig:Kekul_vortex}. 
Note that a disc geometry was chosen to minimize spurious states appearing at the system edges (for instance at corners).
We calculate the eigenstates of the system
${\mathcal H} \ket{\psi_{\alpha}} = \eps_{\alpha} \ket{\psi_{\alpha}}$,
with $\lambda = 1$, and find the LDOS using the formula
\be
{\nu}_{\mu}(E,{\bf r}) = \sum_{\alpha} \delta(E - \eps_{\alpha}) |\bra{0}c_{\mu}({\bf r}) \ket{\psi_{\alpha}}|^2
\; ,
\ee
where $\delta(x)$ is in the ideal case is a Dirac delta function, but for a calculation in a finite system must be taken as some approximation for the delta function. We take a Lorentzian $\delta(x) = \frac{w/\pi}{x^2 + w^2}$ of width $w = 0.00001$
as our approximation.

\begin{figure}[htb]
\centering
\subfigure[$\eps=1.5$]{
\includegraphics[scale=0.3]{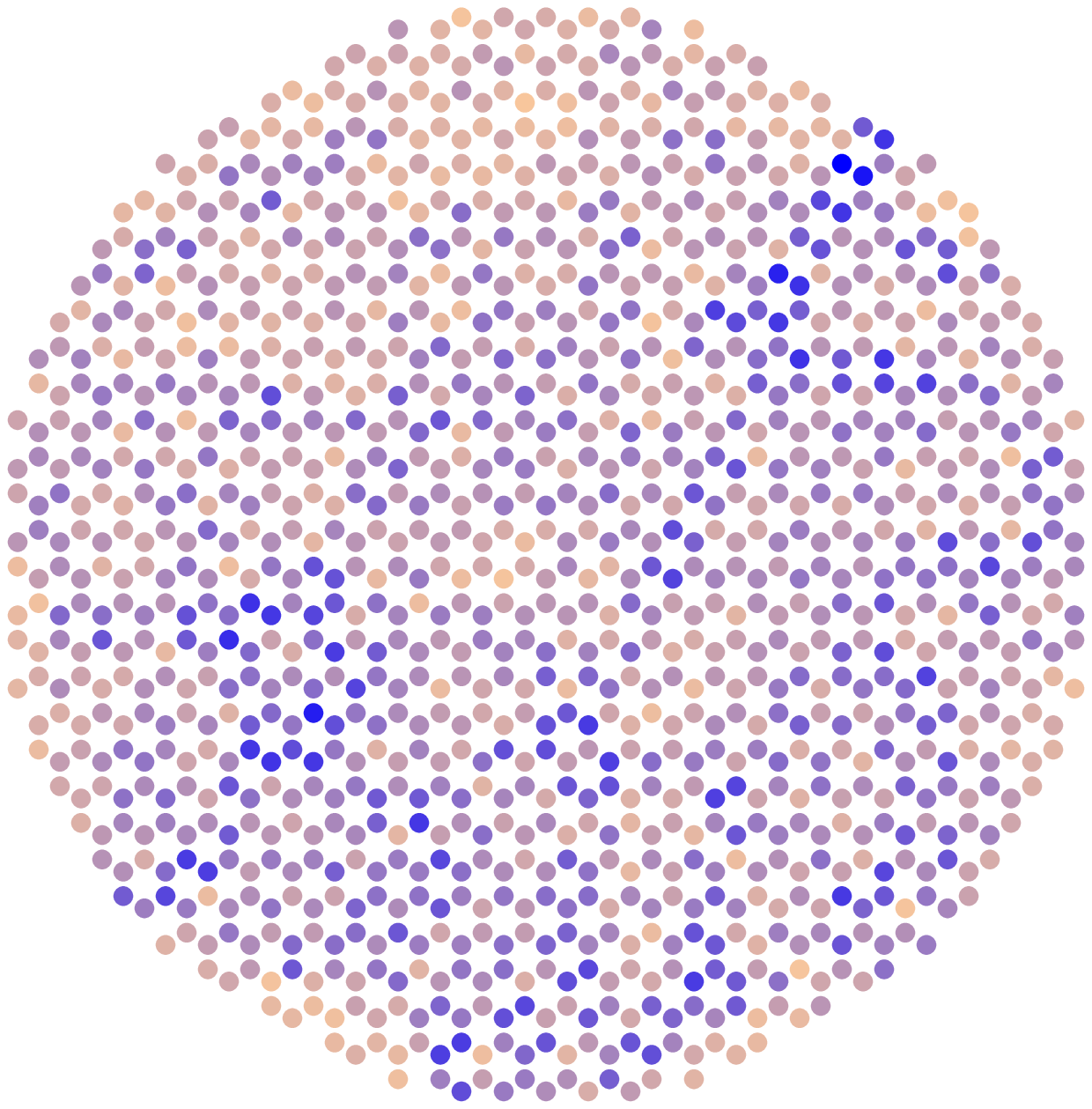}
\label{fig:E15}
}
\subfigure[$\eps=0.5$]{
\includegraphics[scale=0.3]{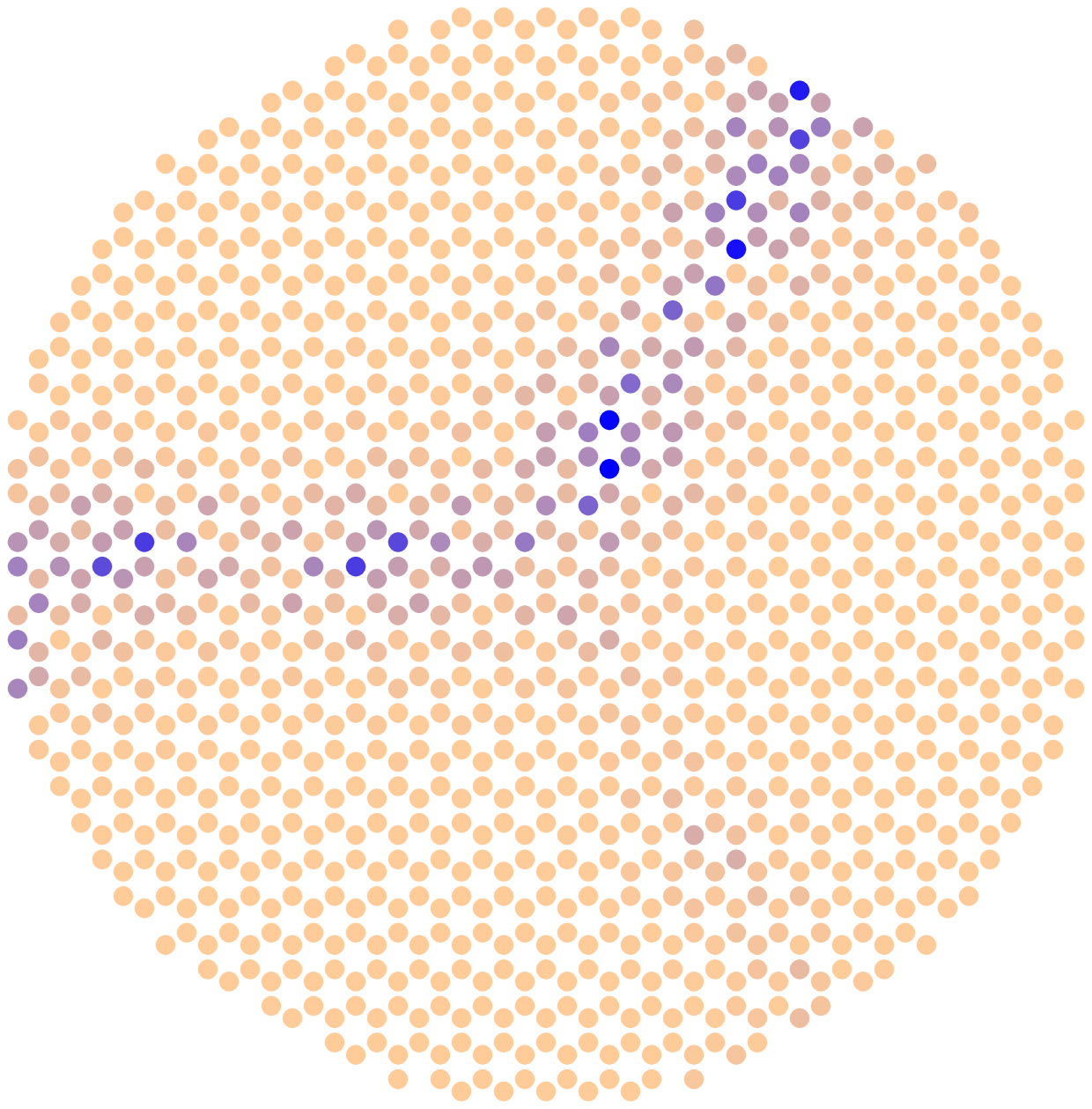}
\label{fig:E05}
}
\subfigure[$\eps=0.1$]{
\includegraphics[scale=0.3]{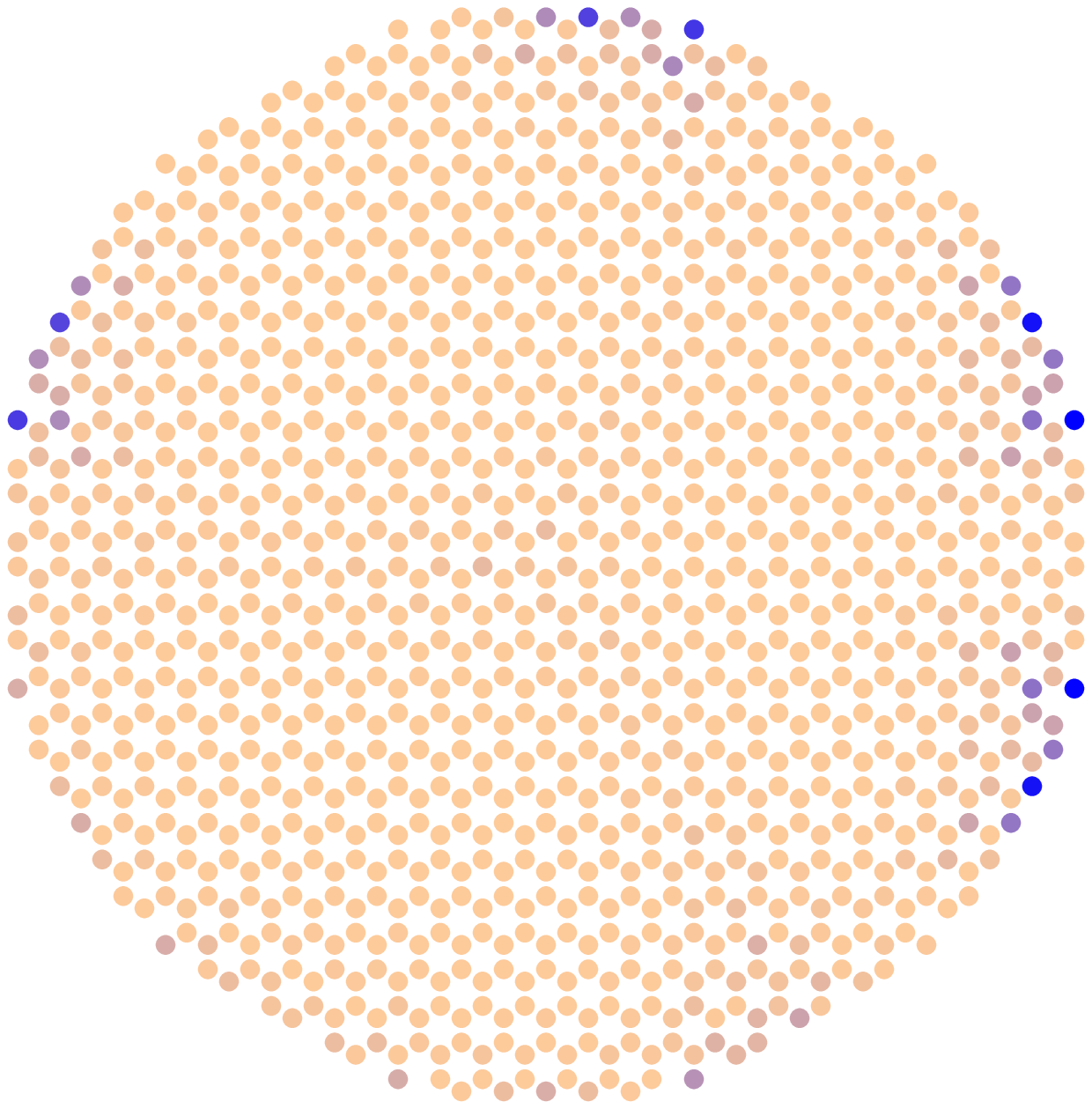}
\label{fig:E01}
}
\subfigure[$\eps=0$]{
\includegraphics[scale=0.3]{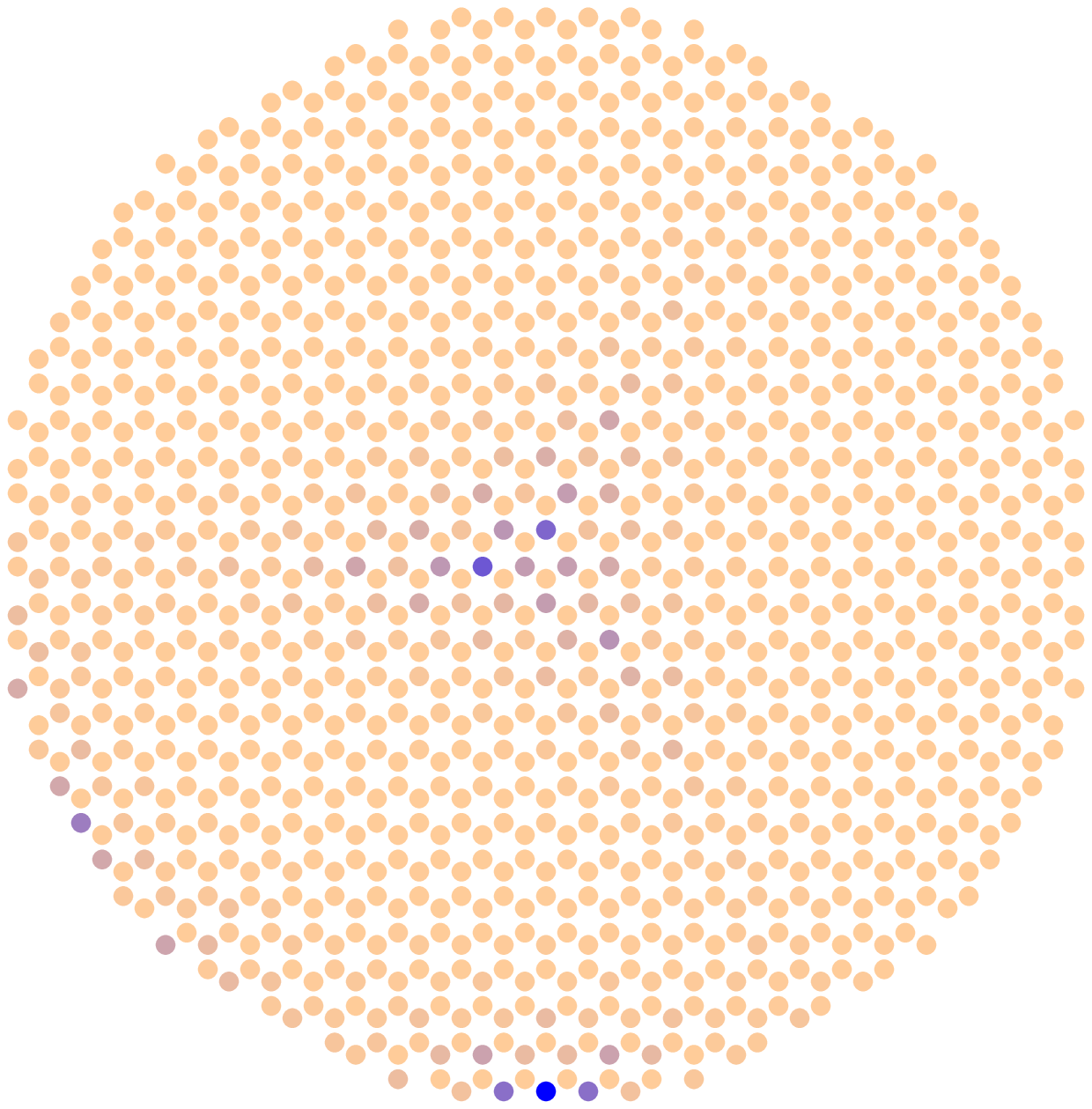}
\label{fig:E00}
}
\subfigure[$\eps=-0.1$]{
\includegraphics[scale=0.3]{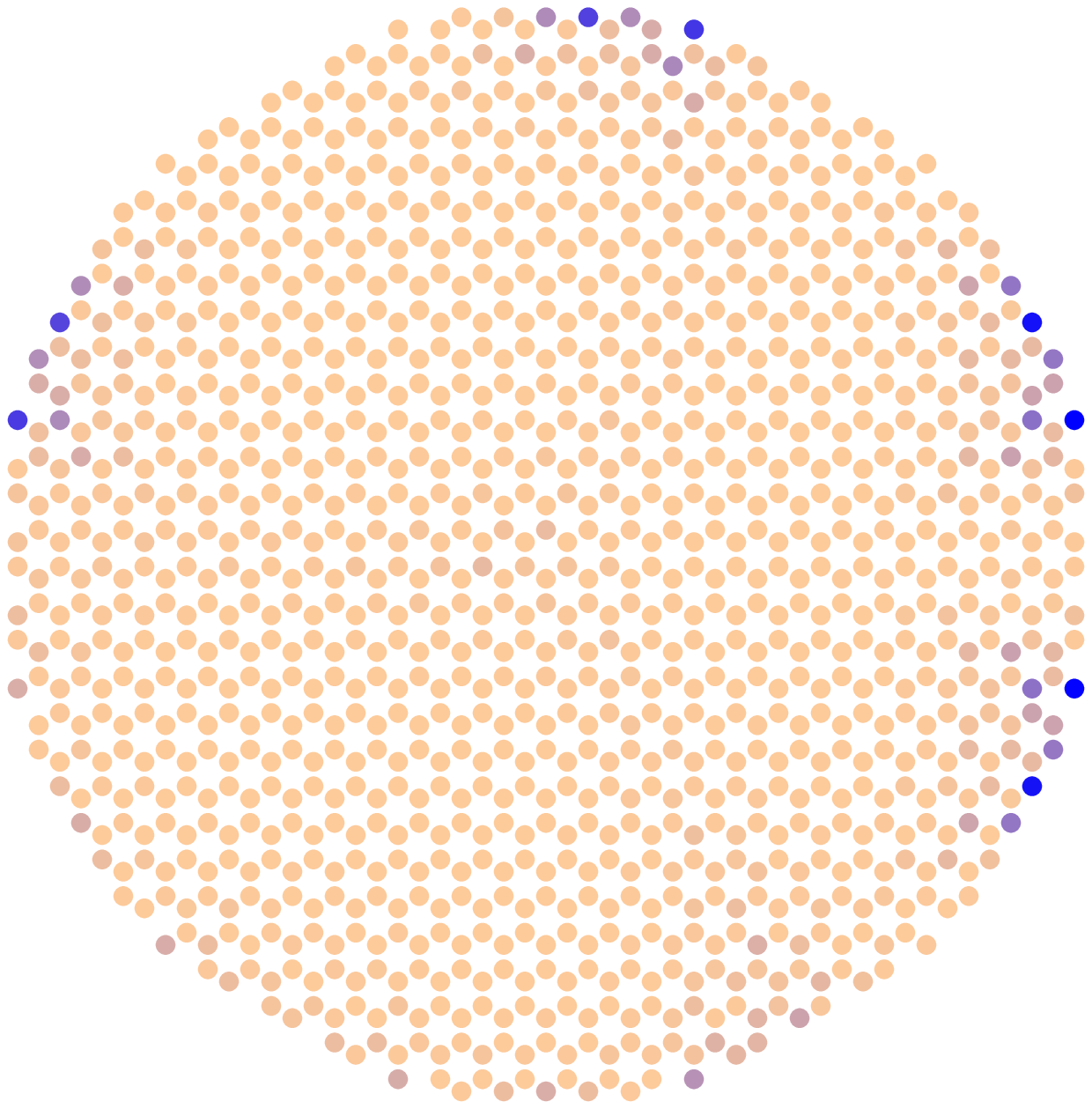}
\label{fig:Em01}
}
\subfigure[$\eps=-0.5$]{
\includegraphics[scale=0.3]{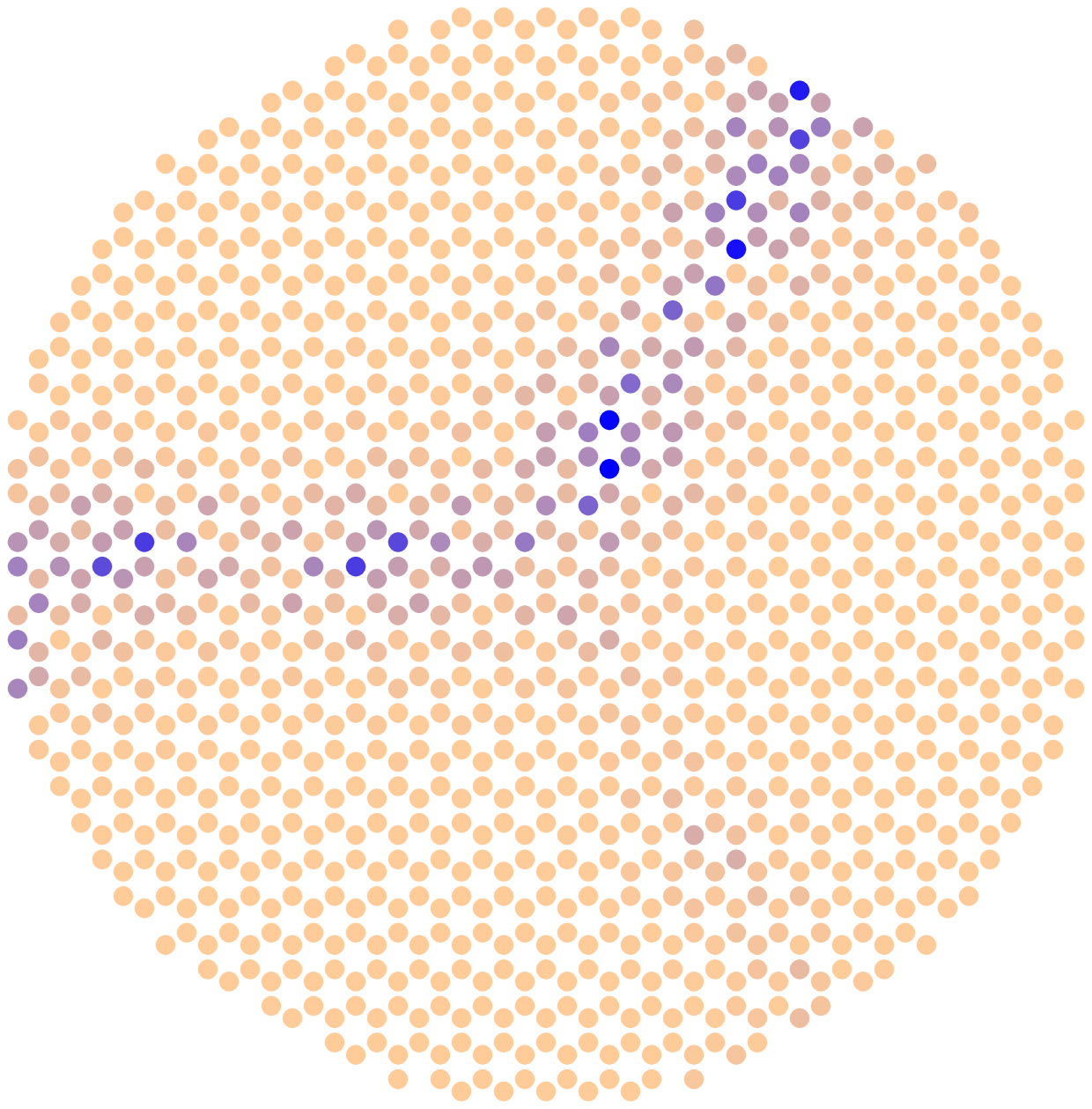}
\label{fig:Em05}
}
\subfigure[$\eps=-1.5$]{
\includegraphics[scale=0.3]{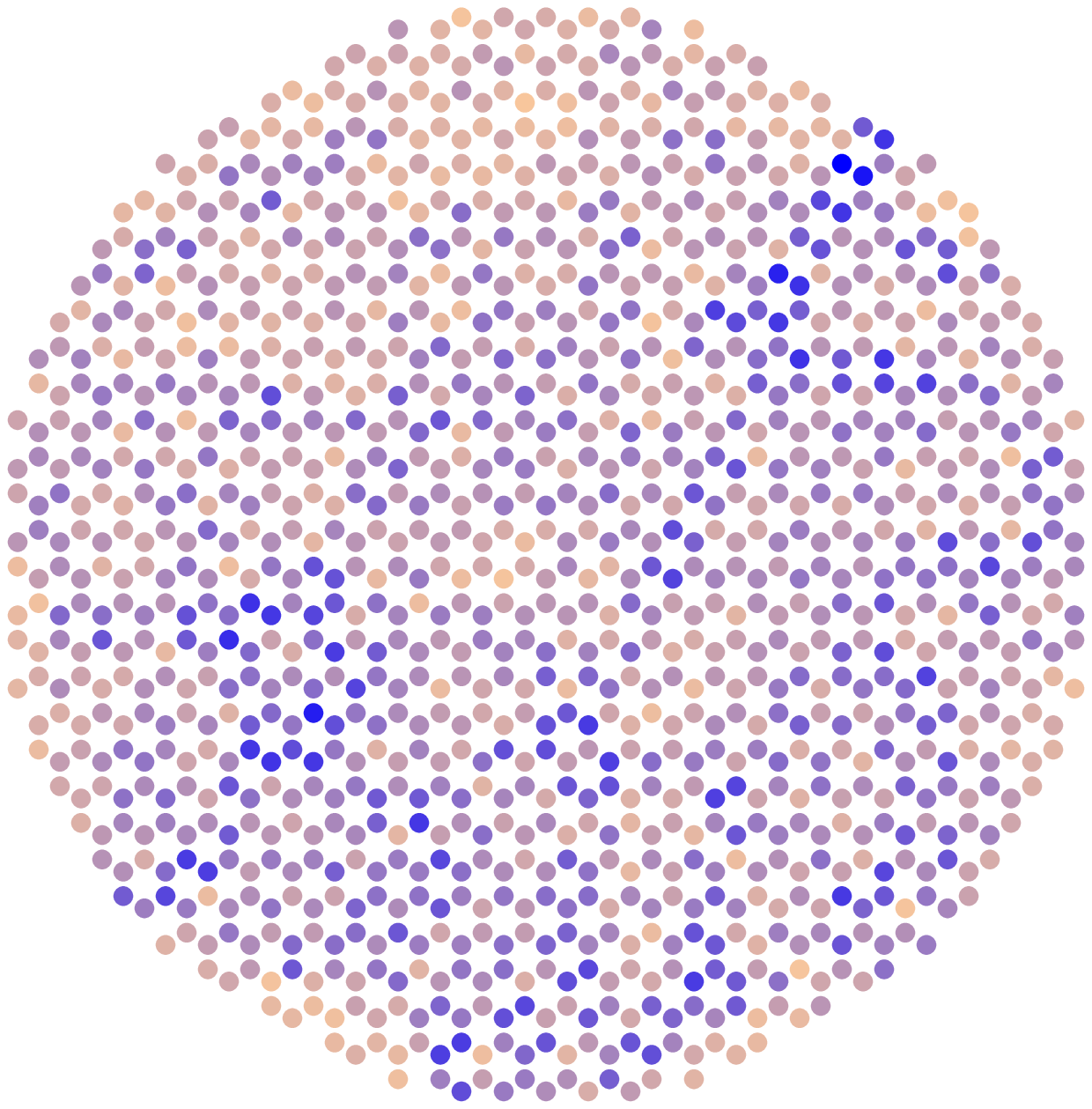}
\label{fig:Em15}
}
\caption{The LDOS for $\lambda = 1$, at different energies: 
$\eps=+1.5$ \subref{fig:E15},
$\eps=+0.5$ \subref{fig:E05},
$\eps=+0.1$ \subref{fig:E01},
$\eps=0$ \subref{fig:E00},
$\eps=-0.1$ \subref{fig:Em01},
$\eps=-0.5$ \subref{fig:Em05}, and
$\eps=-1.5$ \subref{fig:Em15}.
The circles represent sites of the honeycomb lattice, taken here in the shape of a disc of radius $22.5$, using the conventions in the main text. The site coloring is such that dark (blue) points have a higher weight, and lighter (orange) points have lower weight.
While LDOS scans above \subref{fig:E15} and below \subref{fig:Em15} the gap show a rather uniform distribution of DOS, in the gap there clearly is some spatial structure. In particular, at $\eps = 0$ we find a peak at the vortex core, and at one spot on the disc edge \subref{fig:E00}, as expected.}
\label{fig:LDOS}
\end{figure}

We plot the LDOS on the various lattice sites for a number of different energies $E$ in Fig.~\ref{fig:LDOS}, and find that for $E=0$ the LDOS is strongly peaked at the vortex center, and in a spot at the edge of the system (see Fig.~\ref{fig:E00}, each realizing (roughly) one half of an electron. This is precisely what one expects in the case of a halved fermion in a finite system.
In the ideal case, a zero mode appears bound to the vortex core, and another zero mode appears bound to the system edge. These states are degenerate in energy, and any infinitesimal matrix element between them will cause them to form symmetric and anti-symmetric linear combinations, slightly split in energy. The lower of these two energy states is a fermion state delocalized between two positions, with half its wavefunction weight at each spot, regardless of the distance between them.

\begin{figure}[htb]
\centering
\subfigure[]{
\includegraphics[scale=0.3]{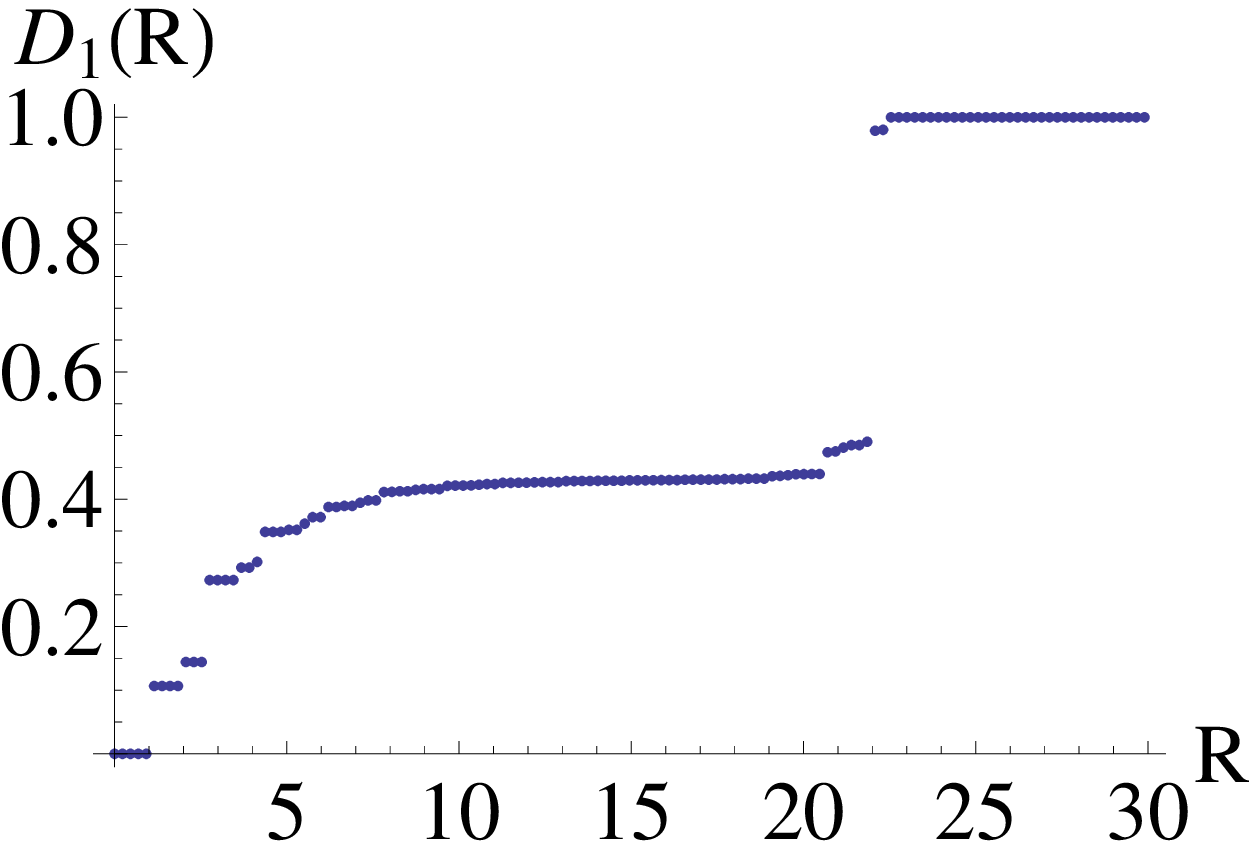}
\label{fig:wavefunction_radial_profile}
}
\subfigure[]{
\includegraphics[scale=0.3]{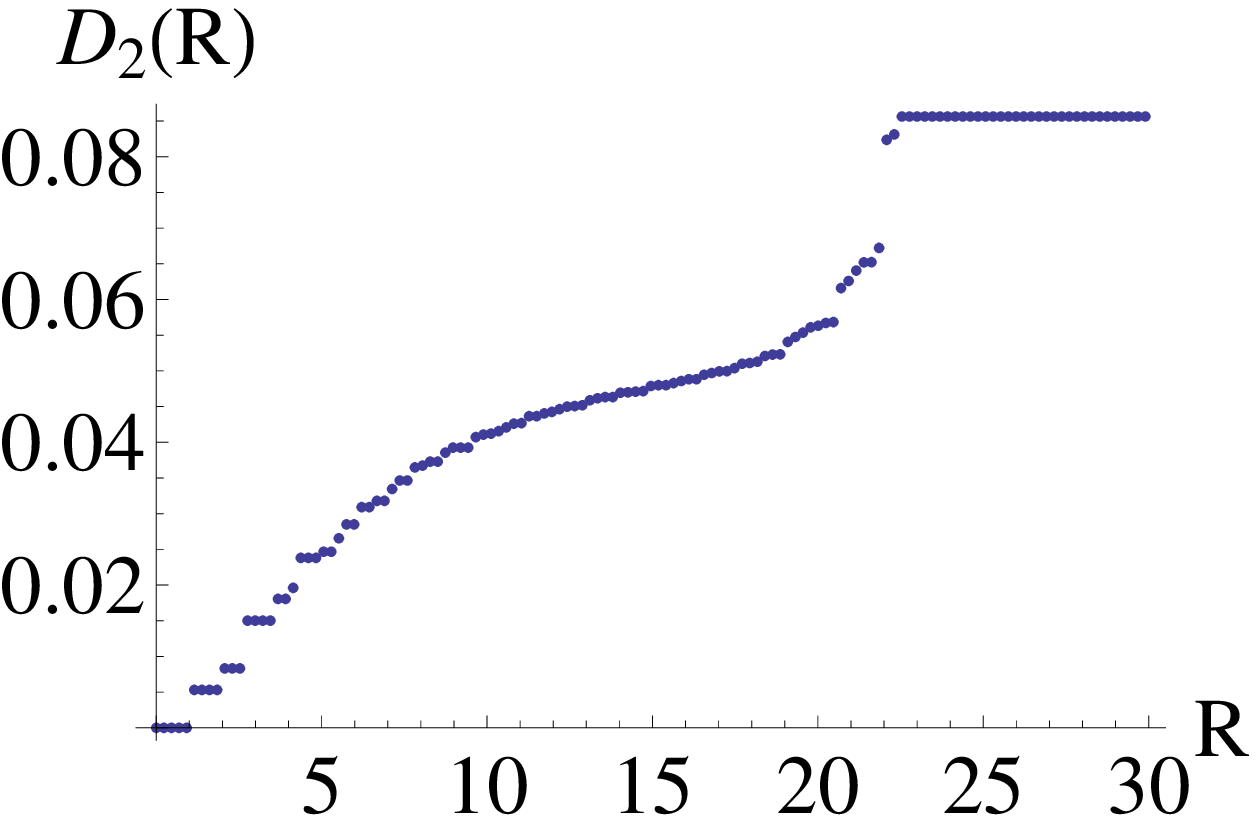}
\label{fig:LDOS_radial_profile}
}
\subfigure[]{
\includegraphics[scale=0.3]{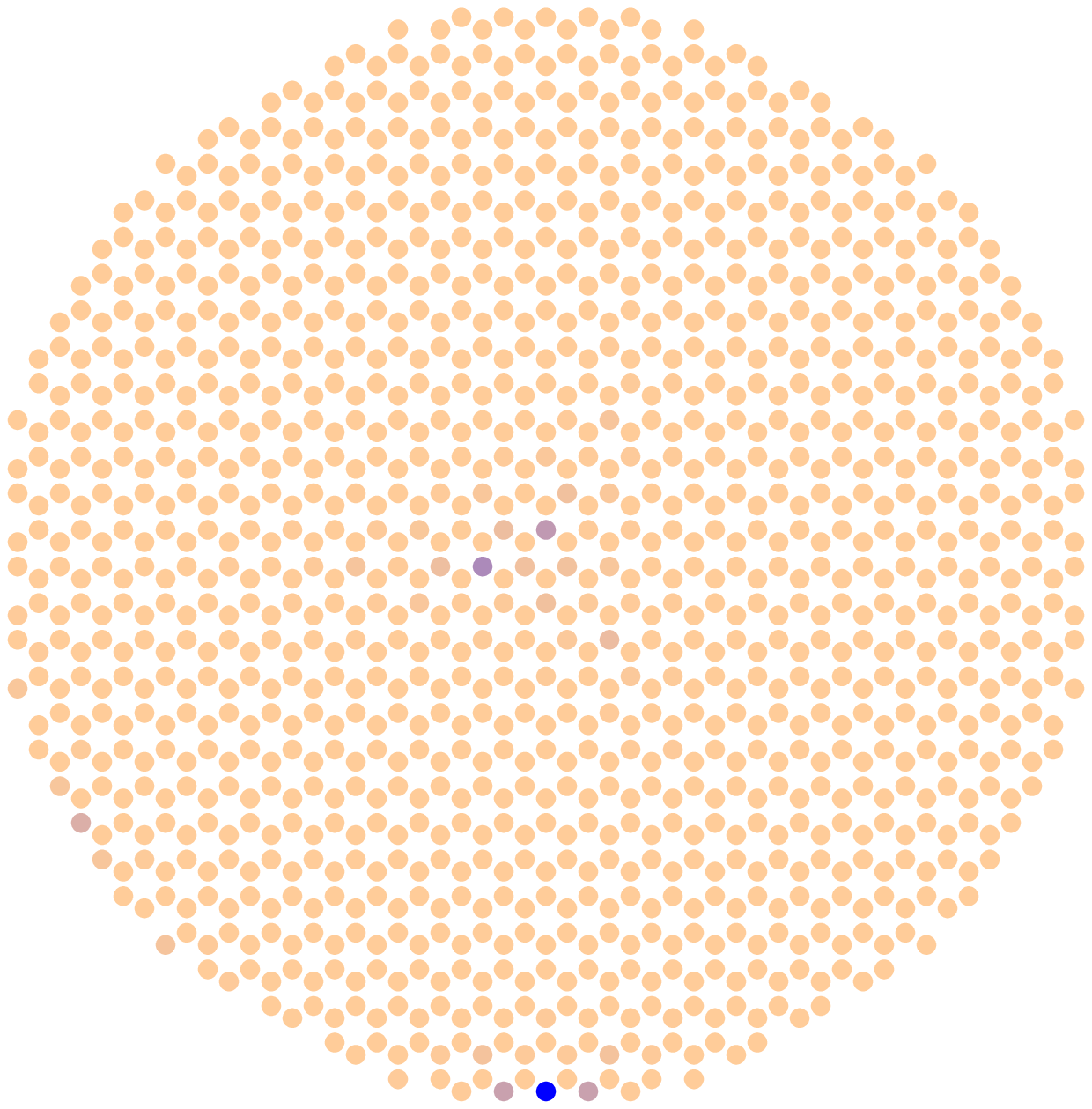}
\label{fig:wavefunc}
}
\caption{Radial accumulated weight of the zero mode wavefunction $D_1(R)$ \subref{fig:wavefunction_radial_profile}, and of the LDOS at $E=0$ $D_2(R)$ \subref{fig:LDOS_radial_profile}. The probability density of the $\psi_{0-}$ wavefunction is depicted in \subref{fig:wavefunc}, using the same convention as for the LDOS plots in Fig.~\ref{fig:LDOS}.}
\label{fig:weight_by_radius}
\end{figure}

Next we will want to find numerically the charge accumulated at the vortex core.
We do this in two ways. Taking the lower energy $E \approx 0$ state,we can integrate its wavefunction weight up to some radius $R$ away from the vortex core
\be
D_1(R) = \int_0^R r dr \int_{-\pi}^{+\pi} d\phi |\psi_{0-}({\bf r})|^2 = 
\sum_{|{\bf r}_j|<R} |\psi_{0-}({\bf r_j})|^2
\; ,
\ee
where $r,\phi$ are the polar coordinates in the plane.
Also, we can take the LDOS at $E=0$, and integrate it up to $R$
\be
D_2(R) = \int_0^R r dr \int_{-\pi}^{+\pi} d\phi \, \nu({\bf r},E=0) = 
\sum_{|{\bf r}_j|<R} \nu({\bf r}_j,E=0)
\; .
\ee
It is important to note that the latter is a quantity we have experimental access to.
We plot the calculated $D_{1,2}(R)$ versus $R$ in Fig.~\ref{fig:weight_by_radius}.
Ideally, only the zero modes at the vortex core and the edge will contribute to the LDOS at $E = 0$.
In a finite system, we will get the sum of the contributions from the two slightly split linear combinations of the zero modes  
\be
\nu({\bf r},E=0) \approx \int_{0-}^{0+} dE \, \nu({\bf r},E) = |\psi_{0-}({\bf r})|^2 + |\psi_{0+}({\bf r})|^2
\; .
\ee
It is therefore expected that half the weight of this quantity will be at the vortex core, and the other half at the edge.
While $D_1(R)$ saturates at a value of $1$, $D_2(R)$ saturates at an arbitrary value, and indicates that we will not get a quantitative measure of the accumulated charge in the vortex core from the LDOS. However, we can still learn a great deal from $D_2$, the quantity STM can measure in the lab, as it does demonstrate that roughly half the overall weight is accumulated in the vortex core, the remaining weight being concentrated near the disc edge. Observing roughly half the total weight centered at the vortex core would suggest that a fractionalized state exists, but this is not conclusive. 
%Observing the opposite, would strongly suggest the absence of a fractionalized state.

Experimental measurement is further complicated by the fact that the electron spin needs to be taken into account. In the carbon monoxide on copper system for molecular Graphene, spin orbit coupling is negligible, and interactions seem to be weak, and therefore all electronic states ought to be spin degenerate. we will still have electron halving between a vortex core and the sample edge, but this will occur for both spin polarizations.
The LDOS measured by STM would be the sum of the contributions from the two spins, but we should still observe a curve like that of Fig.~\ref{fig:LDOS_radial_profile}.
Furthermore, we can move the spin up and down states in opposite direction by applying a Zeeman field, sufficiently weak not to cross any other electronic state, but sufficiently strong to split the different spin zero mode states sufficiently to be observed in the LDOS measurement.

An additional complication arises from the fact that the real system also has a non vanishing second neighbor hopping $t'$ which lowers the symmetry of the Hamiltonian \eqref{Hamiltonian},
and ruins the theoretically perfect $e/2$ fractionalization\cite{Hou:prl2007,Chamon:prl2008}, changing it to some more general fraction. However, the qualitative distinction of the vortex core bound states remains - one electron is delocalized between the vortex core and the edge of the system, with some finite fraction of its weight bound to the vortex core, and the the rest to the edge.

Perhaps a better method than merely measuring the static LDOS, averaged over long times, would be to probe some correlation between the edge and the vortex core, or better yet between two vortex cores. As explained earlier, the effective fermion halving is essentially the delocalizing of a single fermion wavefunction between two spots (for instance two vortex cores). Qualitatively, if the electron is detected near one of the vortex cores at some short time interval, then the wavefunction collapses onto that vortex core, 
and no electron should be detected at the other vortex core during the same short time interval. An experiment probing this temporal correlation could perhaps reveal the fundamental quantum mechanical effect at play here. One could try to simultaneously measure \emph{time resolved} electron tunneling at two locations. Calculating the noise correlation between the two tunneling currents $I_{1,2}$, averaged over time
$\langle \Delta I_1 \Delta I_2 \rangle = \langle I_1 I_2 \rangle - \langle I_1 \rangle \langle I_2 \rangle$, should reveal a long range correlation, only between the vortex cores.
Detecting this would be a strong indicator that a wavefunction at $E\approx0$ is delocalized between these two locations, thus realizing the fermion halving scenario.

%Hwoever such a setup would also have to meet stringent condictions

%\section{finalities}

In conclusion we have demonstrated that the molecular Graphene system can be made to form a Kekule texture with a vortex, thus realizing a physical system with fermion halving. In this case the electron effectively fractionalizes to states with charge $e/2$ bound to the vortex core. The electron spin is expected to merely double the electronic spectrum, and thus the vortex core should accumulate a unit charge, but no magnetization due to spin (see also Ref.~\onlinecite{Hou:prl2007}). The Kekule texture Y--junction has already been realized experimentally\cite{Manoharan:2012}, and it now remains to prove that a fermion halving is indeed occurring in this system. We explored how signatures of the halving would appear in the LDOS, and hope our insights will be tested in the molecular Graphene system.

\emph{Acknowledgments:} DLB was supported by the
Sherman Fairchild foundation, and acknowledges 
the hospitality of the Freiburg Institute of Advanced Studies (FRIAS)
where this work was conceived, and carried out.
Finally, we wish to thank Hari Manoharan, for readily 
discussing his experiments.

\bibliographystyle{apsrev}

\end{document}